\documentclass[aps,prl,reprint,showpacs,byrevtex]{revtex4-2}
\let\oldhat\hat
\renewcommand{\hat}[1]{\oldhat{\mathbf{#1}}}

\usepackage{times,mathptm}
\usepackage[dvips]{graphicx,color}

\begin{document}
\title{Charge density wave and superconductivity in the kagome metal CsV$_3$Sb$_5$ around a pressure-induced quantum critical point}
\author{Chongze Wang$^{1}$, Shuyuan Liu$^1$, Hyunsoo Jeon$^1$, Yu Jia$^{2,3}$, and Jun-Hyung Cho$^{1*}$}
\affiliation{$^1$Department of Physics and Research Institute for Natural Science, Hanyang University, 222 Wangsimni-ro, Seongdong-Ku, Seoul 04763, Republic of Korea \\
$^2$Key Laboratory for Special Functional Materials of the Ministry of Education, Henan University, Kaifeng 475004, People's Republic of China \\
$^3$The Joint Center for Theoretical Physics, Henan University, Kaifeng 475004, China}
\begin{abstract}
Using first-principles density functional theory calculations, we investigate the pressure-induced quantum phase transition (QPT) from the charge density wave (CDW) to the pristine phase in the layered kagome metal CsV$_3$Sb$_5$ consisting of three-atom-thick Sb$-$V$_3$Sb$-$Sb and one-atom-thick Cs layers. The CDW structure having the formation of trimeric and hexameric V atoms with buckled Sb honeycomb layers features an increase in the lattice parameter along the $c$ axis, compared to its counterpart pristine structure having the ideal V$_3$Sb kagome and planar Sb honeycomb layers. Consequently, as pressure increases, the relatively smaller volume of the pristine phase contributes to reducing the enthalpy difference between the CDW and pristine phases, yielding a pressure-induced QPT at a critical pressure $P_c$ of ${\sim}$2 GPa. Furthermore, we find that (i) the superconducting transition temperature $T_c$ increases around $P_c$ due to a phonon softening associated with the periodic lattice distortion of V trimers and hexamers and that (ii) above $P_c$, optical phonon modes are hardened with increasing pressure, leading to monotonous decreases in the electron-phonon coupling constant and $T_c$. Our findings not only demonstrate that the uniaxial strain along the $c$ axis plays an important role in the QPT observed in CsV$_3$Sb$_5$, but also provide an explanation for the observed superconductivity around $P_c$ in terms of a phonon-mediated superconducting mechanism.
\end{abstract}
\maketitle

\section{I. INTRODUCTION}


Recently, the kagome metals AV$_3$Sb$_5$ (A = K, Rb, Cs) have attracted tremendous attention because of their intriguing electronic properties involving nontrivial band topology, charge density wave (CDW) order, and superconductivity (SC). These kagome metals consisting of three-atom-thick Sb$-$V$_3$Sb$-$Sb and one-atom-thick A layers [see Fig. 1(a)] have an interesting electronic structure characterized by the existence of the saddle points of linearly dispersive Dirac bands close to the Fermi level $E_F$. Specifically, they were experimentally observed to exhibit the CDW order at temperatures below about 80$-$100 K~\cite{KV3Sb5-chrialCDW-Nat.Mat2021, CsV3Sb5-SC_CDW-PRL2021, CsV3Sb5-CDW_SC-NC2021, CsV3Sb5-2x2x2CDW-PRX}. This CDW order has been revealed to be unconventional with a chiral anisotropy breaking time-reversal symmetry~\cite{KV3Sb5-chrialCDW-Nat.Mat2021}, which leads to a large anomalous Hall effect in the absence of long-range magnetic ordering~\cite{KV3Sb5-AHC-AS2020, CsV3Sb5-AHC_SC-PRB2021}. However, the origin of CDW is currently being debated whether it is ascribed to the Fermi surface nesting-driven Peierls-like electronic instability~\cite{KagomeFS-PRL2013, AV3Sb5-B.Yan-PRL2021, KagommeCDW-PRB2013, CsV3Sb5-CDW_origin-PRB2021, AV3Sb5-PRB2021, KagomevHs-PRB2021, AV3Sb5_Denner, AV3Sb5_Morten,CsV3Sb5-phonon,CsV3Sb5_Kang_ARPES,CsV3Sb5_Hu_ARPES}, many-body correlations and excitonic effects with particle-hole condensation~\cite{AV3Sb5-CDW-phon_nomaly}, momentum-dependent electron-phonon coupling (EPC)~\cite{KV3Sb5-CDW_Gap, RbV3Sb5-CDW_Gap, CsV3Sb5-CDW_Gap1, CsV3Sb5-CDW_Gap2,CsV3Sb5_natcomm}, or Jahn-Teller-like distortion with the formation of quasimolecular states~\cite{AV3Sb5_chongze}. Upon cooling down to temperatures below ${\sim}$3 K, SC~\cite{CsV3Sb5-Z2-PRL2020,KV3Sb5_SC_Z2_PRM2021,RbV3Sb5-SC-CPL2021} emerges, indicating the coexistence of CDW and SC. Therefore, the AV$_3$Sb$_5$ metals provide an ideal platform to investigate the interplay of CDW and SC on the kagome lattice. Using scanning tunnelling microscope (STM), scanning tunnelling spectroscopy (STS), and Josephson scanning tunnelling spectroscopy, Chen $et$ $al$.~\cite{CsV3Sb5_Roton} observed the V-shaped SC gap on the surfaces of CsV$_3$Sb$_5$ with nonzero local density of states at $E_F$, claiming an unconventional SC coupled to the pair-density wave state. By contrast, another STM and STS study of Xu $et$ $al$.~\cite{CsV3Sb5_Multi} reported a s-wave superconducting pairing symmetry that involves the multiband SC with band-dependent gap distributions, suggesting a conventional Bardeen-Cooper-Schrieffer (BCS) type SC. Therefore, the nature of SC in AV$_3$Sb$_5$ is also still controversial.

The intimate relationship between the intertwined CDW order and SC in AV$_3$Sb$_5$ has been further explored by applying pressure~\cite{CsV3Sb5-CDW_SC-NC2021, CsV3Sb5-SC_CDW-PRL2021, KV3Sb5_pressure_Du, RbV3Sb5_pressure_Wang, CsV3Sb5_SC_PRB2021, CsV3Sb5_HighPress_CPL,CsV3Sb5_HighPress_PRL,CsV3Sb5_Nodal_SC}. The observed pressure$-$temperature ($P-T$) phase diagram displays a quantum phase transition (QPT) between the CDW and pristine phases at zero temperature, which overlaps with a dome-shaped superconducting region~\cite{CsV3Sb5-CDW_SC-NC2021,CsV3Sb5-SC_CDW-PRL2021,KV3Sb5_pressure_Du, RbV3Sb5_pressure_Wang}. For CsV$_3$Sb$_5$, it was experimentally~\cite{CsV3Sb5-CDW_SC-NC2021, CsV3Sb5-SC_CDW-PRL2021} and theoretically~\cite{AV3Sb5_chongze,CsV3Sb5_DFT_Si,CsV3Sb5_DFT_Zhang,CsV3Sb5_Role_of_Sb} reported that the CDW order is suppressed with increasing pressure and transforms into the pristine phase at a critical pressure $P_c$ ${\approx}$ 2 GPa. Around this quantum critical point (QCP), $T_c$ was measured to reach a maximum value of ${\sim}$8 K and then was continuously suppressed to disappear around ${\sim}$15 GPa~\cite{CsV3Sb5-CDW_SC-NC2021, CsV3Sb5-SC_CDW-PRL2021,CsV3Sb5_HighPress_CPL,CsV3Sb5_HighPress_PRL,CsV3Sb5_Nodal_SC}. This superconducting dome is designated as the SC-I region, while there is another superconducting dome (designated as the SC-II region) between 15 and 50 GPa~\cite{CsV3Sb5_HighPress_CPL,CsV3Sb5_HighPress_PRL,CsV3Sb5_Nodal_SC}. The $P-T$ phase diagram in the SC-I region resembles those of many correlated electron systems such as heavy fermion $f$-electron compounds~\cite{CeCu2Si2_Science}, high $T_c$ superconducting cuprates~\cite{Rev-cuprate-2006,Rev-cuprate-2015}, and Fe-based superconductors~\cite{Rev-pnictide-2011,Rev-pnictide-2015}, where quantum fluctuations associated with magnetic or nematic quantum states are the essential ingredients for the SC around the QCP. In this regard, CDW fluctuations around the QCP have been speculated for the emergence of a pressure-induced superconducting dome in CsV$_3$Sb$_5$~\cite{CsV3Sb5-CDW_SC-NC2021, CsV3Sb5-SC_CDW-PRL2021,CsV3Sb5_HighPress_CPL,CsV3Sb5_HighPress_PRL,CsV3Sb5_Nodal_SC}. Meanwhile, several first-principles calculations~\cite{CsV3Sb5_DFT_Si,CsV3Sb5_DFT_Zhang} for CsV$_3$Sb$_5$ showed that the EPC plays an important role in the variation of SC with respect to pressure, supporting a phonon-mediated BCS type mechanism of SC. Thus, the microscopic interplay between the CDW order and SC around the QCP needs to be elucidated.

\begin{figure}[h!t]
\includegraphics[width=8.5cm]{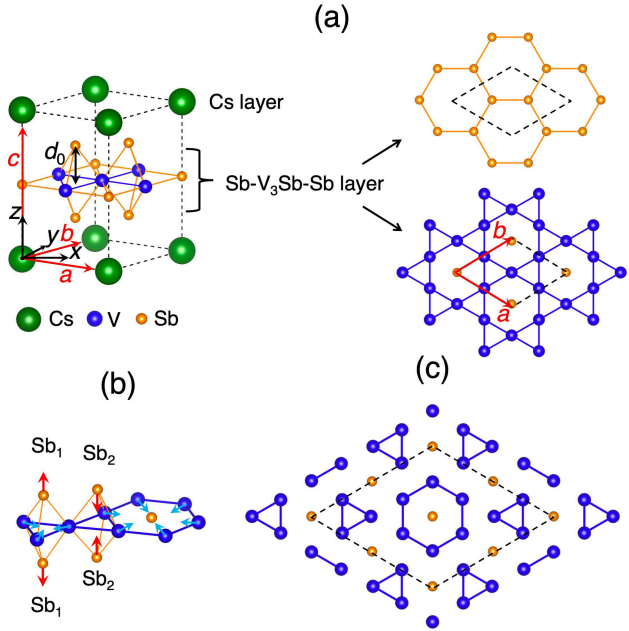}
\caption{Optimized structure of the pristine phase of CsV$_3$Sb$_5$, consisting of the Cs triangular, Sb honeycomb, and V$_3$Sb kagome layers. Here, $d_0$ is the vertical distance between the Sb honeycomb and V$_3$Sb kagome layers. In (b), the arrows represent the lattice distortions of V and Sb atoms in the Sb$-$V$_3$Sb$-$Sb layers, where Sb$_1$ (Sb$_2$) moves along the $c$ axis outward (toward) the kagome plane. The top view of the V$_3$Sb kagome layer in the CDW phase is drawn in (c).The dashed lines in (a) and (c) represent the 1${\times}$1 and 2${\times}$2 unit cells in the $ab$ plane, respectively.}
\label{figure:2}
\end{figure}

In this article, we present first-principles calculations of the structural, electronic, phononic, and superconducting properties of CsV$_3$Sb$_5$ around the QCP in the SC-I region. We find that at the QCP, the CDW phase undergoes a volume contraction to transform into the pristine phase. During this QPT, the formation of trimeric and hexameric V atoms with the buckling of Sb honeycomb layers vanishes in the CDW phase, giving rise to a discontinuous decrease in the lattice parameter along the $c$ axis. Consequently, as pressure increases, the relatively smaller volume of the pristine phase contributes to decrease the magnitude of enthalpy difference ${\Delta}H$ (= ${\Delta}E$ + $P{\Delta}V$) between the two phases, leading to a pressure-induced QPT. It is thus likely that the different uniaxial strain along the $c$ axis between the CDW and pristine phases plays an important role in the QPT of CsV$_3$Sb$_5$. Furthermore, we reveal that, as pressure approaches $P_c$ from higher pressure, the phonon modes associated with the periodic lattice distortion of V trimers and hexamers are softened to yield an elongated lattice parameter along the $c$ axis with buckling Sb honeycomb layers. The resulting enhanced EPC gives rise to an increases in $T_c$ around the QCP. Meanwhile, as pressure increases beyond the QCP, the Cs, V, and Sb-derived optical phonon modes are hardened due to the compressive strain shortening bond lengths, thereby yielding monotonous decreases in the EPC constant and $T_c$. Therefore, the SC around $P_c$ is well explained by a phonon-mediated pairing mechanism.

\section{II. CALCULATIONAL METHODS}

We performed first-principles calculations within the density functional theory (DFT) as implemented in the Vienna ab initio simulation package (VASP) codes~\cite{vasp1,vasp2}. The potential of the core was described by the projector augmented wave method~\cite{paw} and the valence electrons were treated with Cs 3$s^2$3$p^6$4$s^1$, V 3$p^6$3$d^4$4$s^1$, and Sb 5$s^2$5$p^3$ electrons. For the exchange-correlation interaction between the valence electrons, we employed the generalized-gradient approximation functional of Perdew-Burke-Ernzerhof~\cite{pbe}. The van der Waals interaction was included using the DFT-D3 scheme~\cite{DFT-D3-zero}. We used the plane wave basis with a kinetic energy cutoff of 500 eV and performed the $k$-space integration using 12${\times}$12${\times}$8, 6${\times}$6${\times}$8, and 6${\times}$6${\times}$4 $k$-meshes for the 1${\times}$1${\times}$1, 2${\times}$2${\times}$1, and 2${\times}$2${\times}$2 unit cells, respectively. All atoms were allowed to relax along the calculated forces until all the residual force components were less than 0.001 eV/{\AA}. The spectral function on the Fermi surface was computed using Wannier90~\cite{wannnier90} and WannierTools~\cite{wanniertools}, where the maximally localized Wannier functions were constructed by projecting the Bloch states onto V $d$ and Sb $p$ atomic orbitals. The electronic susceptibility was calculated using the EPW code~\cite{EPW} with 120${\times}$120${\times}$80 $k$ points. The phonon spectrum for the pristine phase was calculated using the QUANTUM ESPRESSO package~\cite{qe} with 12${\times}$12${\times}$8 $k$ and 4${\times}$4${\times}$2 $q$ points. For the calculation of EPC, we used the EPW code~\cite{EPW}, with the 48${\times}$48${\times}$32 $k$-meshes and 24${\times}$24${\times}$16 $q$-meshes.

\section{III. RESULTS AND DISCUSSION}

{\bf {A. Pressure-induced QPT between the CDW and pristine phases in CsV$_3$Sb$_5$}} \\

We first optimize the structures of the pristine and CDW phases of CsV$_3$Sb$_5$ as a function of pressure using DFT calculations. The pristine phase crystallizes in the hexagonal space group $P6$/$mmm$ (No. 191) with the stacking of one-atom-thick Cs triangular layer and three-atom-thick Sb$-$V$_3$Sb$-$Sb layers where Sb atoms in the upper and lower Sb honeycomb layers are covalently bonded to V atoms in the middle V$_3$Sb kagome layer containing a triangular Sb sublattice centered on each V hexagon [see Fig. 1(a)]. Meanwhile, the CDW phase was experimentally~\cite{KV3Sb5-chrialCDW-Nat.Mat2021, CsV3Sb5-2x2x2CDW-PRX, CsV3Sb5-SC_CDW-PRL2021, CsV3Sb5-SC_CDW-PRL2021, CsV3Sb5-CDW_SC-NC2021} and theoretically~\cite{AV3Sb5-B.Yan-PRL2021,CsV3Sb5-phonon} revealed to have the so-called tri-hexagonal (TrH) structure where two trimers and one hexamer of V atoms are formed within a 2${\times}$2 unit cell of the V$_3$Sb kagome layer [see Figs. 1(b) and 1(c)]. Hereafter, we consider the 2${\times}$2${\times}$1 TrH structure that captures the key features of the CDW order in the 2D kagome lattice, as discussed below. Figure 2(a) shows the lattice parameter differences, ${\Delta}a$ = ${\Delta}b$ and ${\Delta}c$, between the CDW and pristine structures as a function of pressure. The ${\Delta}a$ and ${\Delta}b$ values are nearly zero with respect to pressure, indicating that the CDW formation hardly influences the lattice parameters along the $a$ and $b$ axes. However, the large positive value of ${\Delta}c$ indicates that the CDW order undergoes a significant tensile strain along the $c$ axis. Consequently, during the QPT between the CDW and pristine phases, the volume contraction occurs in the pristine phase. Therefore, as shown in the inset of Fig. 2(a), there is a discontinuous change ${\Delta}V$ in volume at the QCP of $P_c$ ${\approx}$ 2.1 GPa. We will demonstrate later that the relatively smaller volume of the pristine phase compared to the CDW phase plays an important role in inducing a pressure-induced QPT in CsV$_3$Sb$_5$.

\begin{figure}[h!t]
\includegraphics[width=8.5cm]{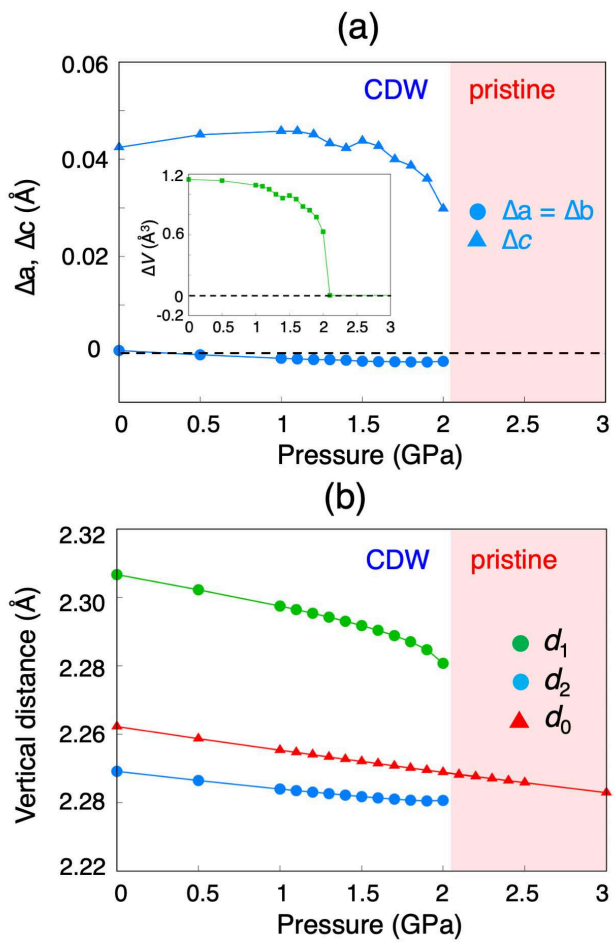}
\caption{(a) Calculated lattice parameter differences between the CDW and pristine structures (i.e., ${\Delta}a$ = $a_{\rm CDW}$ $-$ $a_{\rm pristine}$ and ${\Delta}c$ = $c_{\rm CDW}$ $-$ $c_{\rm pristine}$) as a function of pressure. The inset shows the corresponding volume change ${\Delta}V$ between the CDW and pristine structures. In (b), the vertical distance $d_0$ is plotted as a function of pressure. Here, $d_1$ ($d_2$) represents the vertical distance between Sb$_1$ (Sb$_2$) and the V$_3$Sb kagome layer in the CDW structure. In (a) and (b), the region of the pristine phase is illustrated with pink color.}
\label{figure:2}
\end{figure}

To understand the generation of tensile strain along the $c$ axis in the CDW structure, we analyze the lattice distortion of three-atom-thick Sb$-$V$_3$Sb$-$Sb layers. As shown in Figs. 1(b) and 1(c), the CDW formation changes each planer Sb honeycomb layer into a buckled geometry having two different species of Sb atoms (designated as Sb$_1$ and Sb$_2$). This buckling of Sb atoms is correlated to the formation of trimeric and hexameric V atoms in the V$_3$Sb kagome layer: i.e., the shortened V$-$V bond length in the V trimer brings the displacement of Sb$_1$ (Sb$_2$) along the $c$ axis outward (toward) the kagome plane. Figure 2(b) shows the vertical distance $d_1$ ($d_2$) between Sb$_1$ (Sb$_2$) and the V$_3$Sb kagome plane as a function of pressure. We find that the $d_1$ and $d_2$ values decrease monotonously with increasing pressure and their average value at a certain pressure is longer than the corresponding one [$d_0$ in Fig. 1(a)] in the pristine structure, thereby giving rise to a larger volume of the CDW phase compared to the pristine phase. The resulting tensile strain in the CDW structure is expected to affect the interlayer interaction between neighboring V$_3$Sb kagome layers. Indeed, such repulsive interlayer interaction can be relaxed with the larger 2${\times}$2${\times}$2 TrH structure having alternate stackings of Sb$-$V$_3$Sb$-$Sb layers along the $c$ axis (see Fig. S1 in the Supplemental Material~\cite{SM}), thereby being energetically more favored than the 2${\times}$2${\times}$1 one.

\begin{figure}[h!t]
\includegraphics[width=8.5cm]{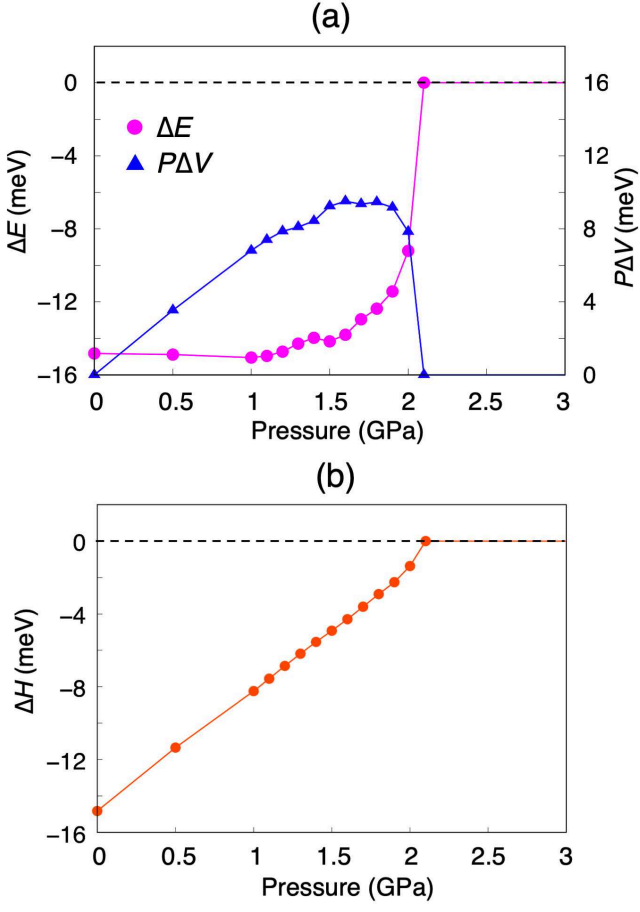}
\caption{(a) Calculated total energy difference ${\Delta}E$ and pressure times volume difference $P{\Delta}V$ between the CDW and pristine phases as a function of pressure. In (b), the corresponding enthalpy difference ${\Delta}H$ is displayed as a function of pressure.}
\label{figure:3}
\end{figure}

The QPT between the CDW and pristine phases at zero temperature is determined by the variation of their enthalpies ($H$ = $E$ + $PV$) as a function of pressure. At pressures below $P_c$, the CDW phase has lower total energy $E$ than the pristine phase. Figure 3(a) shows that ${\Delta}E$ between the CDW and pristine phases changes very little between 0 and 1 GPa, but its magnitude decreases between 1 and 2 GPa. On the other hand, the CDW phase always has larger $PV$ term than the pristine phase because of its relatively larger volume. As shown in Fig. 3(a), $P{\Delta}V$ between the CDW and pristine phases increases linearly between 0 and 1 GPa, but it exhibits a dome shape between 1 and 2 GPa. These variations of negative ${\Delta}E$ and positive $P{\Delta}V$ terms give rise to a monotonous decrease in the magnitude of ${\Delta}H$ with increasing pressure, reaching zero around 2.1 GPa. Thus, the QPT from the CDW to the pristine phase occurs at $P_c$ ${\approx}$ 2.1 GPa, in good agreement with the experimental data~\cite{CsV3Sb5-CDW_SC-NC2021, CsV3Sb5-SC_CDW-PRL2021}. Here, the pressure-induced QPT is accompanied by a discontinuous volume decrease in the pristine phase [see Fig. 2(a)]. It is noticeable that the uniaxial strain along the $c$ axis yielding the volume contraction in the pristine phase plays an important role in the pressure-induced QPT. During this QPT, we find that there are several interesting features in the electronic structure and phonon spectrum of the CDW phase, such as the degeneracy breaking of electronic states around $E_F$ due to the formation of quasimolecular states~\cite{AV3Sb5_chongze} originating from V trimers and hexamers, the resulting reduction of DOS at $E_F$, and the phonon softening associated with the periodic lattice distortion of V trimers and hexamers that are accompanied by the vertical displacements of Sb$_1$ and Sb$_2$ atoms, as will be discussed later. It is thus likely that lattice effects such as Jahn-Teller-like distortion~\cite{AV3Sb5_chongze} and strong EPC~\cite{KV3Sb5-CDW_Gap, RbV3Sb5-CDW_Gap, CsV3Sb5-CDW_Gap1, CsV3Sb5-CDW_Gap2,CsV3Sb5_natcomm} are intimately related to the emergence of the CDW order.

By contrast, many experimental~\cite{CsV3Sb5-CDW_origin-PRB2021,CsV3Sb5_Kang_ARPES,CsV3Sb5_Hu_ARPES} and theoretical~\cite{AV3Sb5-PRB2021,AV3Sb5-B.Yan-PRL2021,KagomevHs-PRB2021} studies of CsV$_3$Sb$_5$ have raised the issue of electron correlation effects for the CDW formation. At ambient pressure, ARPES experiments~\cite{CsV3Sb5_Kang_ARPES,CsV3Sb5_Hu_ARPES} observed several saddle points of linearly dispersive Dirac bands near $E_F$, which are located at the $M$ point in the Brillouin zone. The resulting van Hove singularities (vHs) were proposed to drive the CDW order via Fermi surface nesting~\cite{KagomeFS-PRL2013, KagommeCDW-PRB2013}. To examine how such a Fermi surface nesting mechanism is attributed to the pressure-induced QPT, we calculate the electronic band structure and Fermi surface of the pristine phase at 2.1 GPa. Figures 4(a) and 4(b) show the calculated band structure and four Fermi surface sheets $FS_1$, $FS_2$, $FS_3$, and $FS_4$, respectively. We find that $FS_1$ arises mostly from Sb $p_z$ orbital; $FS_2$ from V $d_{xy}$, $d_{x^2-y^2}$, and $d_{z^2}$ orbitals; $FS_3$ and $FS_4$ from V $d_{xz}$ and $d_{yz}$ orbitals [see Figs. 4(a) and S2]. These orbital characters and overall shapes of $FS_1$, $FS_2$, $FS_3$, and $FS_4$ are similar to the corresponding ones calculated at zero pressure (see Figs. S3 and S4 in the Supplemental Material~\cite{SM}). In order to quantitatively estimate the strength of Fermi surface nesting~\cite{Marin-PRB2008}, we calculate the low-frequency limit of the imaginary part of electronic susceptibility~\cite{Marin-PRB2008}, defined as Im[${\chi_0}$($q$)] $=$ $\sum_{nm} \int d\mathbf{k} \delta(\varepsilon_{n\mathbf{k}}-E_{F}) \delta(\varepsilon_{m\mathbf{k+q}}-E_{F})$. As shown in Fig. 4(c), Im[${\chi_0}$($q$)] obtained at 2.1 GPa exhibits broad peaks around the $M$ point. Since the shape of Fermi surface is insensitive with respect to pressure~\cite{note-FS} and there is no dominant peak of Im[${\chi_0}$($q$)] at the $M$ point, the Fermi surface nesting-driven Peierls-like electronic instability~\cite{KagomeFS-PRL2013, AV3Sb5-B.Yan-PRL2021, KagommeCDW-PRB2013, CsV3Sb5-CDW_origin-PRB2021, AV3Sb5-PRB2021, KagomevHs-PRB2021, AV3Sb5_Denner, AV3Sb5_Morten,CsV3Sb5-phonon,CsV3Sb5_Kang_ARPES,CsV3Sb5_Hu_ARPES} is unlikely the driving force of CDW order.

\begin{figure}[h!t]
\includegraphics[width=8.5cm]{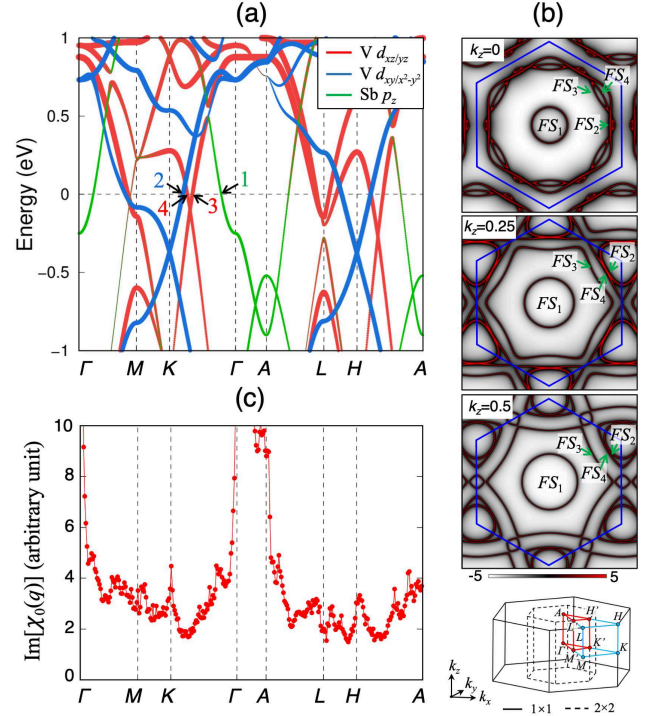}
\caption{(a) Electronic band structure of the pristine phase calculated at 2.1 GPa. Here, the bands are projected onto V $d_{xz}$/$d_{yz}$ (red), V $d_{xy}$/$d_{x^2-y^2}$ (blue), and Sb $p_z$ (green) orbitals, where the radii of circles are proportional to the weights of the corresponding orbitals. The numbers 1, 2, 3, and 4 represent the points of $FS_1$, $FS_2$, $FS_3$, and $FS_4$ crossing $E_F$ along the ${\Gamma}-K$ line, respectively. In (b), four Fermi surface sheets are displayed at $k_z$ = 0, 0.25, and 0.5. The spectral function on each Fermi surface sheet is drawn using the color scale. The Brillouin zones of the 1${\times}$1${\times}$1 pristine and 2${\times}$2${\times}$1 CDW phases are drawn on the bottom. In (c), the low-frequency limit of the imaginary part of electronic susceptibility is plotted along the high symmetry lines.}
\label{figure:4}
\end{figure}

Below the QCP, the periodic lattice distortion of V trimers and hexamers produces the breaking of the degeneracies of electronic states $S_1$, $S_2$, $S_3$, and $S_4$ around $E_F$ [see Fig. 5(a)]. We find that the band dispersion of the CDW phase along the ${\Gamma}-A$ line exhibits the variation of the CDW-induced gaps ${\Delta}_i$ ($i$ = 1, 2, 3, and 4) with respect to pressure, which arise from the $S_1$, $S_2$, $S_3$, and $S_4$ states, respectively. The values of ${\Delta}_i$ obtained at the ${\Gamma}$ point decrease with increasing pressure [see Fig. 5(b)]. These reduced ${\Delta}_i$ values under pressure are consistent with the decrease in the magnitude of ${\Delta}E$ between 1 and 2 GPa [see Fig. 3(a)]. It is noted that ${\Delta}_i$ along the ${\Gamma}-A$ line is spread over the energy range away from $E_F$. This behavior of ${\Delta}_i$ also contradicts the previously proposed Fermi surface nesting mechanism~\cite{AV3Sb5-PRB2021,AV3Sb5-B.Yan-PRL2021,KagomevHs-PRB2021,CsV3Sb5-CDW_origin-PRB2021,CsV3Sb5_Kang_ARPES,CsV3Sb5_Hu_ARPES} that usually opens CDW gaps at the nesting vectors around $E_F$.

\begin{figure}[h!t]
\includegraphics[width=8.5cm]{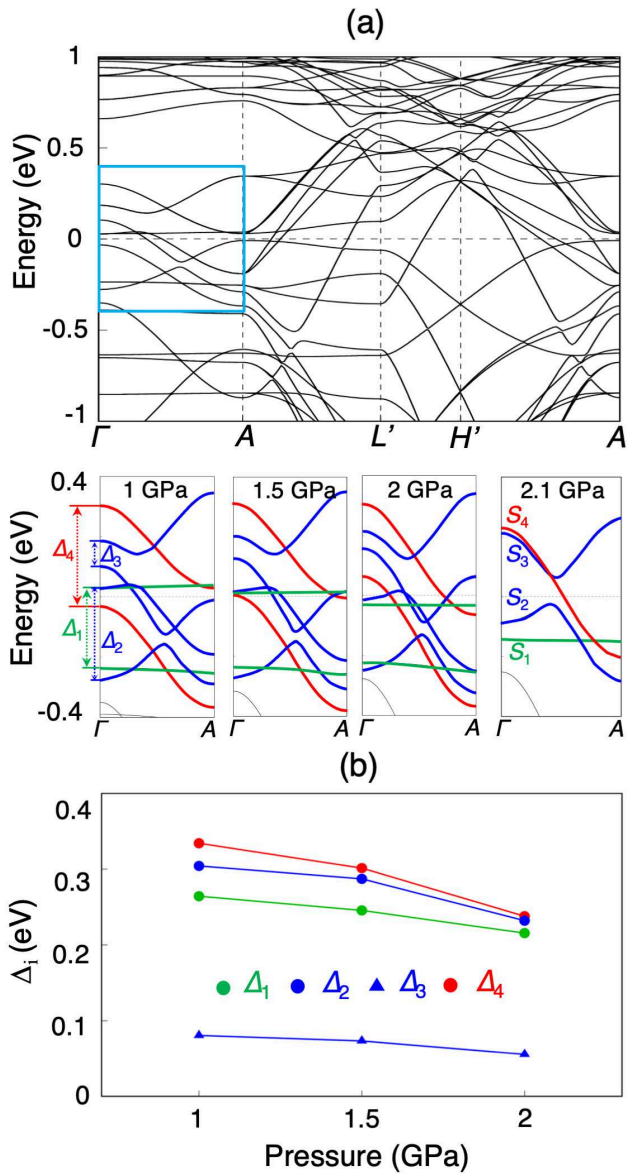}
\caption{(a) Electronic band structures of the CDW phase calculated at 1 GPa, together with the closeup band dispersions of the $S_1$, $S_2$, $S_3$, and $S_4$ states along ${\Gamma}-A$ line. The results obtained at 1.5, 2, and 2.1 GPa are also given. In (b), the CDW gaps ${\Delta}_i$ of the $S_1$, $S_2$, $S_3$, and $S_4$ states are displayed as a function of pressure.}
\label{figure:4}
\end{figure}

So far, we have considered the CDW phase using the 2${\times}$2${\times}$1 TrH structure. We find that at 2.0 GPa, the 2${\times}$2${\times}$2 TrH structure having alternate stackings of Sb$-$V$_3$Sb$-$Sb layers (see Fig. S1 in the Supplemental Material~\cite{SM}) becomes more stable than the 2${\times}$2${\times}$1 one only by ${\sim}$6 meV per pristine unit cell. Here, the band dispersions of the 2${\times}$2${\times}$1 and 2${\times}$2${\times}$2 TrH structures along the ${\Gamma}-M'-K'-{\Gamma}$ line in the $k_x$-$k_y$ plane are similar to each other (see Fig. S5). Moreover, since the predicted value of $P_c$ ${\approx}$ 2.1 GPa for the QCP between the 2${\times}$2${\times}$1 TrH and pristine structures is close to the experimental data~\cite{CsV3Sb5-CDW_SC-NC2021, CsV3Sb5-SC_CDW-PRL2021} of ${\sim}$2 GPa, we believe that the 2${\times}$2${\times}$1 TrH structure captures the major physics of the QPT.

\vspace{0.4cm}
{\bf {B. Superconductivity of compressed CsV$_3$Sb$_5$ around QCP}} \\

To investigate the SC of CsV$_3$Sb$_5$ beyond the QCP, we calculate the phonon spectrum of the pristine phase as a function of pressure using the density functional perturbation theory~\cite{dfpt1,dfpt2}. Figures 6(a), 6(b), and 6(c) show the phonon spectra calculated at 3, 6, and 9 GPa, respectively. From their projected DOS onto Cs, V, and Sb atoms, we find that the optical phonon modes arising from Cs, V, and Sb atoms are somewhat separated with each other: e.g., at 3 GPa, Cs-derived modes are mostly distributed in a frequency range around ${\sim}$60 cm$^{-1}$, Sb-derived modes, between ${\sim}$70 and ${\sim}$170 cm$^{-1}$, and V-derived modes, between ${\sim}$180 and ${\sim}$310 cm$^{-1}$. However, the low-frequency acoustic phonon modes show a strong momentum dependence with respect to the three atoms. It is noticeable that, as pressure increases, the optical phonon modes shift upwards in the overall frequency range [see Figs. 6(a), 6(b), and 6(c)], indicating a pressure-induced phonon hardening. Here, compression makes all the bonds shorter, which in turn renders their associated phonon modes harder. Meanwhile, at 3 GPa, the acoustic modes around the $L$ point are found to be largely softened to lower frequencies [see the arrow in Fig. 6(a)], resulting in imaginary-frequency modes below $P_c$ (see Fig. S6 in the Supplemental Material~\cite{SM}). Such phonon softening gives rise to a periodic lattice distortion of V trimers and hexamers, forming the TrH structure. It is noted that out-of-plane strain is sensitive to the variation of pressure, while in-plane strain is nearly negligible (see Fig. S7). Therefore, as pressure approaches $P_c$ from higher pressure, a prominent strain along the $c$ axis weakens the Sb$-$V bonding between the planar Sb honeycomb and V$_3$Sb kagome layers, and therefore the formation of V trimers and hexamers would be easily enabled by a softening of the in-plane vibrations of V atoms. It is thus likely that the uniaxial strain along the $c$ axis plays an important role in driving the pressure-induced QPT between the pristine and CDW phases.

\begin{figure}[h!t]
\includegraphics[width=8.5cm]{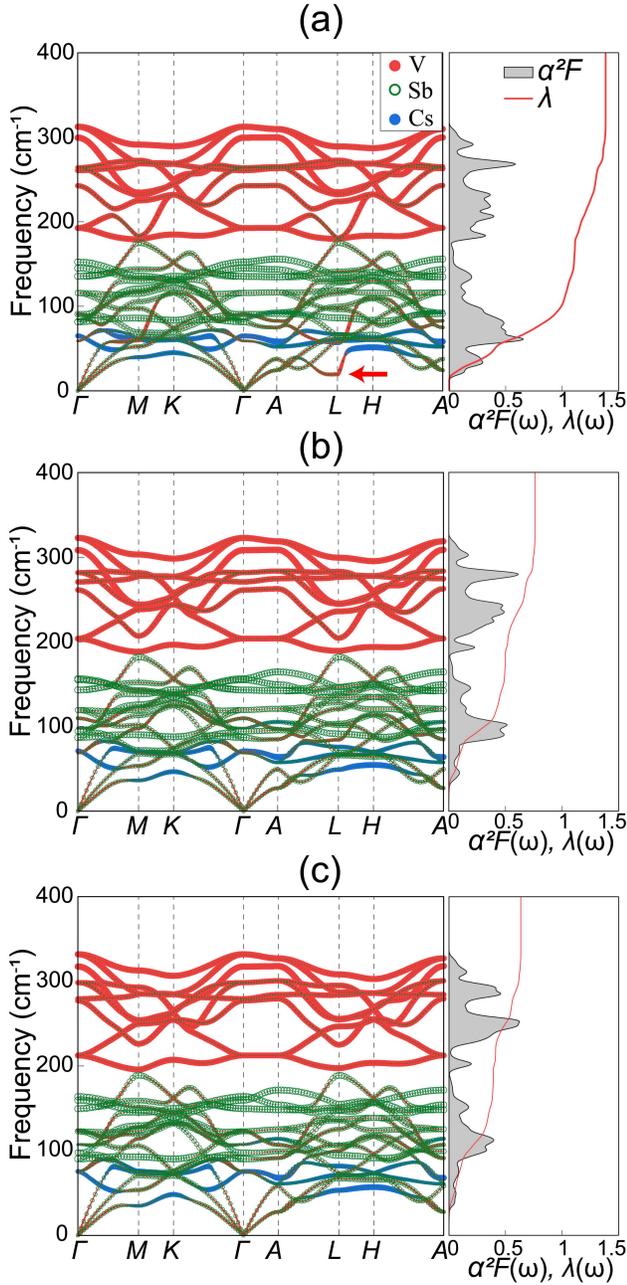}
\caption{Calculated phonon spectra of the pristine phase at (a) 3 GPa, (b) 6 GPa, and (c) 9 GPa. Here, the size of circles on the phonon dispersion is proportional to the weight of the vibrational modes of different atoms. The Eliashberg function ${\alpha}^{2}F({\omega})$ and integrated EPC constant ${\lambda}({\omega})$ are also included as a function of phonon frequency.}
\label{figure:6}
\end{figure}

Next, we calculate the electronic band structure of the pristine phase as a function of pressure. The calculated band structures at 3, 6, and 9 GPa are displayed in Figs. 7(a), 7(b), and 7(c), respectively. Their projected bands onto V 3$d$ and Sb 5$p$ orbitals are also displayed in Fig. S8 in the Supplemental Material~\cite{SM}. We find that the V 3$d$-derived states comprising the $FS_2$, $FS_3$, and $FS_4$ Fermi surface sheets slightly change with respect to pressure, whereas the Sb 5$p_z$-derived states are significantly varied as a function of pressure [see the arrows in Figs. 7(a), 7(b), and 7(c)]. Accordingly, the electron (hole) pocket around the ${\Gamma}$ ($A$) point shifts upward with increasing pressure, leading to a disappearance (appearance) of its corresponding Fermi surface sheet around 6 (9) GPa. The resulting decrease in the DOS at $E_F$ with increasing pressure [see Fig. 7(d)] contributes to lower $T_c$, as discussed below. It is noted that for the pristine phase, the electronic states near $E_F$ are mostly characterized by V 3$d$ and Sb 5$p_z$ orbitals [see Fig. 4(a)]. These V- and Sb-derived states are likely to effectively screen the low-frequency acoustic phonon modes near the QCP, thereby giving rise to the above-mentioned phonon softening around the $L$ point [see Figs. 6(a) and S5].

\begin{figure}[h!t]
\includegraphics[width=8.5cm]{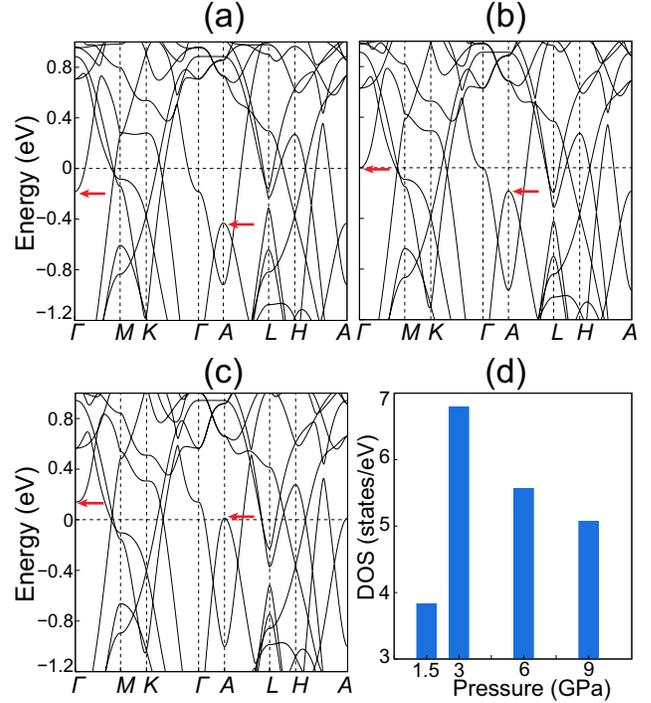}
\caption{Electronic band structures of the pristine phase calculated at (a) 3 GPa, (b) 6 GPa, and (c) 9 GPa. Here, the arrows indicate significant pressure-dependent shifts of the Sb 5$p_z$-derived states at the ${\Gamma}$ and $A$ points. In (d), the DOS (in the unit of states/eV per pristine unit cell) at $E_F$ is displayed as a function of pressure.}
\label{figure:7}
\end{figure}

\begin{figure}[h!t]
\includegraphics[width=8.5cm]{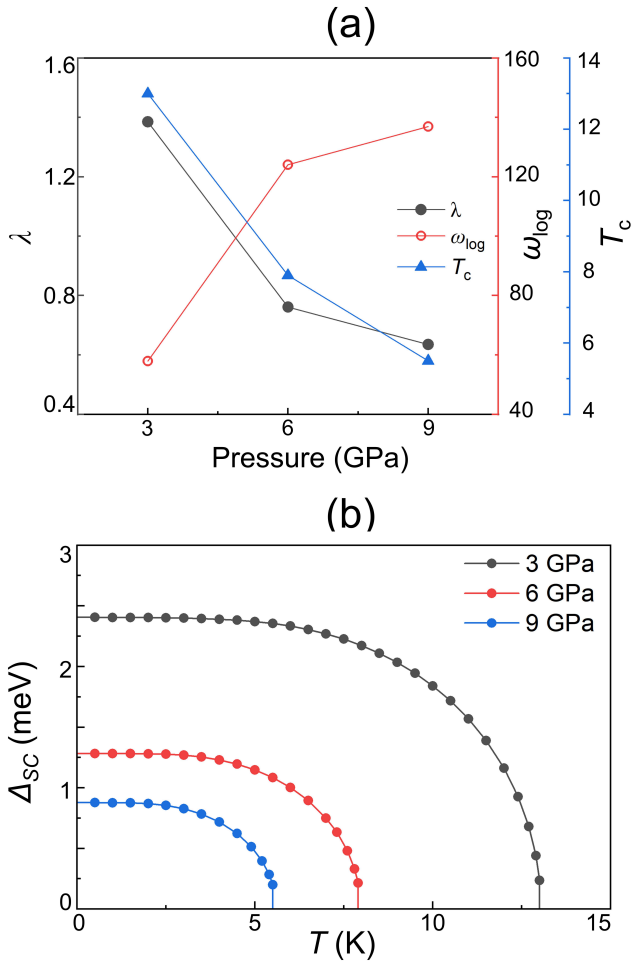}
\caption{(a) Estimated values of ${\lambda}$, ${\omega}_{\rm log}$, and $T_{\rm c}$ as a function of pressure. In (b), the temperature dependence of ${\Delta}_{\rm sc}$ is displayed at different pressures.}
\label{figure:8}
\end{figure}

To estimate the EPC and $T_c$ of the pristine phase under pressure beyond the QCP, we calculate the Eliashberg function ${\alpha}^{2}F({\omega})$ and the integrated EPC constant ${\lambda}({\omega})$ as a function of phonon frequency ${\omega}$ using the isotropic Migdal-Eliashberg equations~\cite{Migdal,Eliash,ME-review}. The results of ${\alpha}^{2}F({\omega})$ and ${\lambda}({\omega})$ obtained at 3, 6, and 9 GPa are displayed in Figs. 6(a), 6(b), and 6(c), respectively. It is seen that ${\alpha}^{2}F({\omega})$ and ${\lambda}({\omega})$ are dominantly contributed from optical phonon modes. For instance, at 3 GPa, the V-, Sb-, and Cs-derived optical phonon modes above a frequency of 50 cm$^{-1}$ contribute to ${\sim}$71 \% of the total EPC constant ${\lambda}$ = ${\lambda}$(${\infty}$). As shown in Fig. 8(a), ${\lambda}$ decreases with increasing pressure as 1.39, 0.76, and 0.64 at 3, 6, and 9 GPa, respectively. Meanwhile, the logarithmically average phonon frequency ${\omega}_{\rm log}$ increases with increasing pressure as 58, 124, and 137 cm$^{-1}$ at 3, 6, and 9 GPa, respectively [see Fig. 8(a)]. By numerically solving the isotropic Eliashberg equations~\cite{Eliash} with the typical Coulomb pseudopotential parameter of ${\mu}^*$ = 0.13~\cite{CsV3Sb5_DFT_Si,CsV3Sb5_DFT_Zhang}, we obtain the temperature dependence of superconducting gap ${\Delta}_{\rm sc}$. As shown in Fig. 8(b), ${\Delta}_{\rm sc}$ closes at $T_c$ ${\approx}$ 13.0, 7.9, and 5.5 K at 3, 6, and 9 GPa, respectively. In view of the Allen-Dyne formula~\cite{ADnote} where the decrease (increase) in ${\lambda}$ (${\omega}_{\rm log}$) contributes to lower (raise) $T_c$, we can say that ${\lambda}$ is more likely responsible for the observed~\cite{CsV3Sb5-CDW_SC-NC2021, CsV3Sb5-SC_CDW-PRL2021,CsV3Sb5_HighPress_CPL,CsV3Sb5_HighPress_PRL,CsV3Sb5_Nodal_SC} lowering of $T_c$ with increasing pressure. Here, the presently predicted $T_c$ values between 3 and 9 GPa are reduced as large as ${\sim}$7.5 K, in good agreement with the experimental data~\cite{CsV3Sb5-CDW_SC-NC2021, CsV3Sb5-SC_CDW-PRL2021,CsV3Sb5_HighPress_CPL,CsV3Sb5_HighPress_PRL,CsV3Sb5_Nodal_SC} showing a decrease of ${\sim}$6 K between 2 and 9 GPa.

It is worth noting that, as pressure approaches $P_c$ from higher pressure, the phonon softening around the $L$ point is enhanced and the DOS at $E_F$ increases [see Fig. 7(d)]. As a result, the estimated values of ${\lambda}$, ${\Delta}_{\rm sc}$($T$), and $T_c$ increase around the QCP, as shown in Figs. 8(a) and 8(b). Meanwhile, below the QCP, the phonon softening around the $L$ point disappears through the CDW formation. Moreover, the TrH structure opens the CDW gaps ${\Delta}_i$ (see Fig. 5), thereby reducing the DOS at $E_F$. As shown in Fig. 7(d), the DOS at $E_F$ obtained at 1.5 GPa is much smaller than that of the pristine structure at 3 GPa. It is therefore natural to expect that the $T_c$ vs pressure relation can exhibit a dome shape around the QCP, consistent with the experimentally observation with a maximum $T_c$ around 2 GPa~\cite{CsV3Sb5-CDW_SC-NC2021, CsV3Sb5-SC_CDW-PRL2021,CsV3Sb5_HighPress_CPL,CsV3Sb5_HighPress_PRL,CsV3Sb5_Nodal_SC}. We note that the pressure-induced structural transition between the CDW and pristine phases accompanies a strong EPC effect via the phonon softening of the pristine structure around the $L$ point. This phonon softening in turn leads to a dome shape of $T_c$ around the QCP. We thus conclude that the SC around the QCP can be explained in terms of a phonon-mediated pairing mechanism.

\section{IV. CONCLUSION}

Our first-principles DFT calculations for CsV$_3$Sb$_5$ have shown that the QPT from the CDW to the pristine phase is accompanied by a volume contraction in the pristine phase. We found that the relatively smaller volume of the pristine phase contributes to reduce the enthalpy difference between the CDW and pristine phases with increasing pressure, leading to a pressure-induced QPT around ${\sim}$2 GPa. It was revealed that the discontinuous volume change at the QCP is due to the structural distortion in three-atom-thick Sb$-$V$_3$Sb$-$Sb layers: i.e., the formation of trimeric and hexameric V atoms with the buckling of Sb honeycomb layers is vanished to yield a discontinuous decrease in the lattice parameter along the $c$ axis. Furthermore, we found that the phonon modes associated with the periodic lattice distortion of V trimers and hexamers are softened around the QCP, thereby increasing the EPC and $T_c$. It is thus likely that the phonon-mediated pairing mechanism is responsible for the SC around the QCP. Therefore, our findings not only demonstrate that the volume contraction of the pristine phase plays an important role in inducing the pressure-induced QPT in CsV$_3$Sb$_5$, but also shed light on the phonon-mediated SC around the QCP.

\vspace{0.4cm}

\noindent {\bf Acknowledgements.}
This work was supported by the National Research Foundation of Korea (NRF) grant funded by the Korean Government (Grant No. 2022R1A2C1005456), by BrainLink program funded by the Ministry of Science and ICT through the National Research Foundation of Korea (Grant No. 2022H1D3A3A01077468), and by the National Natural Science Foundation of China (Grant No. 12074099). The calculations were performed by the KISTI Supercomputing Center through the Strategic Support Program (Program No. KSC-2022-CRE-0073) for the supercomputing application research.  \\


\begin{thebibliography}{99}
\bibitem{note-FS} If Fermi surface nesting were the driving force of CDW order, the Fermi surface should significantly change with respect to pressure because the CDW formation is suppressued with increasing pressure.
\end{thebibliography}
\end{document}